\documentclass[journal,12pt,onecolumn]{IEEEtran}

\usepackage[english]{babel}
\usepackage{amsmath,amssymb,amscd,latexsym,dsfont}
\usepackage{float,color,graphicx,subfigure,balance}
\usepackage{multirow,multicol}
\usepackage{comment,cite}
\usepackage{enumerate}
\usepackage{stfloats}
\usepackage{psfrag}
\usepackage{cite}

\IEEEoverridecommandlockouts

\newcommand{\tr}[1]{\textrm{#1}}
\newcommand{\mb}[1]{\mathbf{#1}}
\newcommand{\ov}[1]{\overline{#1}}

\newcommand{\mc}[1]{\mathcal{#1}}

\newcommand{\set}[1]{\{#1\}}

\newcommand{\cd}{\cdot}
\newcommand{\ld}{\ldots}

\begin{document}

\title{Outage Probability of Diversity Combining Receivers in Arbitrarily Fading Channels}
\author{%
\authorblockN{Mohammed Jabi, Leszek Szczecinski$^*$, and Mustapha Benjillali\\}
\authorblockA{INPT, Rabat, Morocco\\}
\authorblockA{$^*$ INRS-EMT, Montreal, Canada\\}

{\normalsize [\emph{jabi.mohamed@gmail.com}, \emph{leszek@emt.inrs.ca}, \emph{benjillali@ieee.org}]}

\thanks{The work was supported by the government of Quebec, under grant \#PSR-SIIRI-435. When this work was submitted for publication M.~Jabi was at INRS, Montreal, Canada.}
}%

\maketitle
\thispagestyle{empty}

\begin{abstract}
We propose a simple and accurate method to evaluate the outage probability at the output of arbitrarily fading $L$-branch diversity combining receiver. The method is based on the  saddlepoint approximation, which only requires the knowledge of the moment generating functions of the signal-to-noise ratio at the output of each diversity branch. In addition, we show that the obtained results reduce to closed-form expressions in many particular cases of practical interest. Numerical results illustrate a very high accuracy of the proposed method for practical outage values and for a large mixture of fading and system parameters.
\end{abstract}

\begin{IEEEkeywords}
\begin{center}
Cooperative Communications, Diversity Combining, HARQ, Hoyt fading, Nakagami-$m$ fading,\\ Outage Probability, Rice Fading, Saddlepoint Approximation.
\end{center}
\end{IEEEkeywords}

\section{Introduction}\label{Sec:Introduction}
\IEEEPARstart{I}{n this} work, we propose a simple method to evaluate the outage probability at the receiver after combining independently fading signals whose distributions are known. Outage analysis is a fundamental problem in communications theory, and has been extensively studied in the literature. It consists in finding the cumulative distribution function (CDF) of the sum of independent random variables. Exact expressions (e.g., \cite{Alouini01,Suraweera06,Sagias08,Benjillali10c}) or closed-form approximations \cite{daCosta08} can be obtained in particular cases and, in more general situations, the problem can be solved using a numerical integration (via the inverse Laplace transform) \cite{Ko00}. 

The approach we propose is valid when fading distributions are arbitrary but with known moment generating functions (MGFs). We propose to use the so-called {\it saddlepoint approximation} (SPA) \cite{Butler07_book}, a well known tool of statistical analysis that was already applied to solve various problems in the area of communications \cite{Helstrom90,Mckay06,Shin06,Martinez07,Du07,Kenarsari10}.

Using SPA, we are able to approximate---very accurately---the outage probability of diversity combining receivers over arbitrarily fading channels. The problem at this level of generality was solved in \cite{Ko00} using Laplace transform. Here, unlike \cite{Ko00},  we are able to provide closed-form solutions in two particularly interesting cases: (i) when combining $L$ identically distributed signals (corresponding to Nakagami-$m$, Rice, or Hoyt fading), and (ii) when combining $L=2$ non-identically distributed Nakagami-$m$ fading signals. Our closed-form approximations of the outage probability are accurate and very simple, while calculating the exact expressions requires numerical integration, as only in particular cases, it is known in closed-form.

\section{System Model}\label{Model}

We consider a communication system where the same signal is received on $L$ independent \emph{diversity branches} at the receiver. While the  term \emph{branches} evokes the receiver connected to different front-ends (which reflects the combining implementation in multi-antenna receivers), the scenario we consider  also covers relay-based communications and automatic repeat request (ARQ) retransmissions. In these cases the $l$-th diversity branch will thus correspond, respectively, to the $l$-th relay's transmission and the $l$-th ARQ transmission round. 

We assume that the channel between the transmitter and the receiver is varying (fading) randomly from one transmission to another but stays invariant during each of the transmissions (block-fading channel), thus the signal received on the $l$-th diversity branch is given by 
\begin{align}
  \mb{y}_{l}=\sqrt{\gamma_{l}}\cd \mb{x} + \mb{z}_{l},
\end{align}
where $\mb{z}_{l}$ is a zero-mean, unitary variance Gaussian signal modelling noise, $\mb{x}$ is the unitary-variance transmitted signal, and $\gamma_{l}$ is the instantaneous signal-to-noise ratio (SNR) on the $l$-th branch. We assume that $\gamma_{l}$ is a random variable whose probability density function (PDF) is given by $\tr{p}_{\gamma_{l}}(\gamma)$ and the corresponding MGF -- by $\mc{M}_{l}(s)=\mathbb{E}_{\gamma_{l}}\!\left[\tr{e}^{\gamma_{l}\cd s}\right]$, where  $\mathbb{E}_{x}[\cd]$ denotes the mathematical expectation calculated with respect to $x$. The average SNR is denoted by $\ov{\gamma}_{l}=\mathbb{E}_{\gamma_{l}}[\gamma_{l}]$. In this work, we consider three of the most common fading distributions, namely, Nakagami-$m$, Rice, and Hoyt. Table~\ref{Table:CGF} shows the corresponding MGFs.

\begin{table}[t]
\centering
\begin{tabular}{l||c|c|c}

Fading type & Nakagami-$m$ & Rice & Hoyt\\ \hline\hline

Parameters  & $m_{l}$  &  $K_{l}$, $K'_{l}=K_{l}+1$  & $q_{l}$, $q'_{l}={(q_{l}+1)}^{2} $\\ \hline

$\mc{M}_{l}(s)$  & $\displaystyle{\left(\frac{m_{l}}{m_{l}-s\ov{\gamma_{l}}}\right)^{\!m_{l}}}$ &  $\displaystyle{\frac{K'_{l}}{K'_{l}-s\ov{\gamma}_{l}}
\tr{exp}{\left( \frac{K_{l}s\ov{\gamma}_{l}}{K_{l}'-s\ov{\gamma}_{l}} \right)}}$  &  $\displaystyle{\left(1-2s\ov{\gamma}_{l}+\frac{q_{l}(2s\ov{\gamma}_{l})^{2}}{q'_{l}}\right)^{\!-0.5}}$\\ \hline

$\kappa'_{l}(s)$  &  $\displaystyle{\frac{m_{l}\ov{\gamma}_{l}}{m_{l}-s\ov{\gamma}_{l}}}$ &  $\displaystyle{\frac{((K'_{l})^{2}-s\ov{\gamma}_{l})\ov{\gamma}_{l}}{(K'_{l}-s\ov{\gamma}_{l})^{2}}}$ &  $\displaystyle{\frac{ q'_{l}\ov{\gamma}_{l} -s q_{l} (2\ov{\gamma}_{l})^{2}} {q'_{l}(1-2s\ov{\gamma}_{l})+{q}_{l}(2s\ov{\gamma}_{l})^{2}}}$  \\ \hline

$\kappa''_{l}(s)$ &  
$\displaystyle{\frac{m_{l}\ov{\gamma}^{2}_{l}}{\left(m_{l}-s\ov{\gamma}_{l}\right)^{2}}}$  &  
$\displaystyle{\frac{(K'_{l}(1+2K_{l})-s\ov{\gamma}_{l})\ov{\gamma}_{l}^{2}}{(K'_{l}-s\ov{\gamma}_{l})^{3}}}$  & 
$\displaystyle{\frac{-(2\ov{\gamma}_{l})^{2}q_{l}} {q'_{l}(1-2s\ov{\gamma}_{l})+{q}_{l}(2s\ov{\gamma}_{l})^{2}} + 2\ov{\gamma}_{l}^{2}\left( \frac{ 4s\ov{\gamma}_{l}q_{l}-q'_{l}} {q'_{l}(1-2s\ov{\gamma}_{l})+{q}_{l}(2s\ov{\gamma}_{l})^{2}} \right)^{2}}$ 
\\ \hline

$\kappa'''_{l}(0)$  &  $\displaystyle{\frac{2\ov{\gamma}^{3}_{l}}{m_{l}^{2}}}$  &
$\displaystyle{\frac{2(1+3K_{l})\ov{\gamma}_{l}^{3}}{(K'_{l})^{3}}}$  &  $-8\displaystyle{\ov{\gamma}_{l}^{3}\left(\frac{3{q}_{l}}{q'_{l}}-1\right)}$

\end{tabular}
\caption{MGF $\mc{M}_{l}(s)$ of the SNR and the derivatives of the corresponding CGF $\kappa_{l}(s)=\log\mc{M}_{l}(s)$ for the adopted fading models.}\label{Table:CGF}
\end{table}

At the receiver, maximum ratio combining (MRC) of the $L$ branches is performed, and the combined signal can be written as
\begin{align}
  \mb{y}=\sum_{n=1}^{L}\sqrt{\gamma_{l}}\mb{y}_{l}=\sqrt{\gamma}\cd\left(\sqrt{\gamma}\cd\mb{x}+\mb{z}\right),
\end{align}
where $\gamma=\sum_{n=1}^{L}\gamma_{l}$ is the aggregate SNR after combining and $\mb{z}$ is a zero-mean unitary variance Gaussian signal modelling the equivalent noise at the output of the combiner.

Let $\gamma_\tr{th}$ be the SNR threshold at the receiver below which the communication is in outage\footnote{This could be seen as the minimum SNR equivalent to an error rate corresponding to a QoS constraint or, in the case of coded transmissions, to the SNR below which successful decoding is not possible}. The outage probability can thus be expressed as
 \begin{align}\label{P.out}
P_\tr{out} = \tr{Pr}\set{ \gamma<\gamma_\tr{th}  }=\tr{F}_{\gamma}(\gamma_\tr{th}),
 \end{align}
where $\tr{F}_{\gamma}(x)=\int_{0}^{x}\tr{p}_{\gamma}(\gamma)\tr{d}\gamma$ is the CDF of $\gamma$. Clearly, finding the outage probability boils down to calculating the CDF $\tr{F}_{\gamma}(x)$. 

Using an infinite-series representation \cite{Alouini01}, the CDF can be obtained for Nakagami-$m$ distribution for arbitrary values of $\ov{\gamma}_{l}$ and shape parameters $m_{l}$, cf.~Table~\ref{Table:CGF}. Also for Nakagami-$m$ with integer parameters $m_l$, closed-form expressions exist in selective diversity combining scenarios \cite{Benjillali10c}.

An alternative approach to evaluate \eqref{P.out} is based on the inverse Laplace transform \cite{Ko00}. Since $\gamma$ is the sum of $L$ independent random variables, the Laplace transform of its PDF (or, in other words, its MGF) is the product of $L$ individual Laplace transforms. As explained in \cite{Ko00}, using the MGFs of the variables, the outage probability in \eqref{P.out} can be found using numerical integration. 

Here, we propose an approach that keeps the generality of the solution in \cite{Ko00} valid for arbitrarily distributed variables $\gamma_{l}$, as long as their MGFs are known. However, our solution  (i)~does not require numerical integration used in \cite{Ko00}, and (ii)~yields closed-form solution for identically distributed $\gamma_{l}$. Moreover, using our approach, a closed-form outage expression is obtained in the case where $L=2$ Nakagami-$m$ distributed signals are combined. The solution is valid for any values of $m_{l}$ and $\ov{\gamma_{l}}$, and it is an alternative to the analytical results shown in \cite{Benjillali10c} which are limited to integer $m_{l}$.

\section{Saddlepoint Approximation of Outage Probability}\label{Sec:Num}

We use here the so-called saddlepoint approximation which is a simple and accurate method to approximate the CDF of a random variable. Knowing $\kappa(s)=\log\mathbb{E}_{\gamma}[\tr{e}^{\gamma s}]$ -- the CGF of $\gamma$, the CDF of $\gamma$ can be approximated by \cite[Ch.~1]{Butler07_book}
\begin{align}\label{SPA}
    \tr{F}_{\gamma}(x)\approx \hat{\tr{F}}_{\gamma}(x)=
\begin{cases}
Q(-\hat{w})+\phi(\hat{w})\cd\displaystyle{\left(\frac{1}{\hat{w}}-\frac{1}{\hat{u}}\right)} & \text{if}\quad x\neq \mathbb{E}_{\gamma}[\gamma]\\
\displaystyle{\frac{1}{2}+ \frac{\kappa'''(0)}{6\sqrt{2\pi}\bigl[\kappa''(0)\bigr]^{3/2}}}& \text{if}\quad x= \mathbb{E}_{\gamma}[\gamma]
\end{cases},
\end{align}
with
\begin{align}
\hat{w}&=\tr{sign}(\hat{s})\sqrt{2\bigl(\hat{s}\cd x-\kappa(\hat{s})\bigr)}, \quad \hat{u}=\hat{s}\sqrt{\kappa''(\hat{s})},\nonumber
\end{align}
where $\kappa'(s)$, $\kappa''(s)$ and $\kappa'''(s)$  are respectively the first, the second and the third derivatives of $\kappa(s)$, $\phi(x)=\displaystyle{\frac{1}{\sqrt{2\pi}}}\exp(-\frac{x^{2}}{2})$, $Q(x)=\displaystyle{\int_{x}^{\infty}\phi(x)\tr{d}x}$, and  $\hat{s}$ depends on $x$ through the saddlepoint equation
\begin{align}
\kappa'(\hat{s})&=x;\label{SP.eq}
\end{align}
solving \eqref{SP.eq} is the most difficult step of the SPA-based approach. 

In our case, $\kappa(\hat{s})=\sum_{l=1}^{L}\kappa_{l}(\hat{s})$ so, once $\hat{s}$ is found, we can use  the relevant expressions for  the $p$-th derivative of the CGF of  $\kappa_{l}(s)$, shown in Table~\ref{Table:CGF},  to calculate $\kappa'(\hat{s})=\sum_{l=1}^{L}\kappa'_{l}(\hat{s})$ and $\kappa''(\hat{s})=\sum_{l=1}^{L}\kappa''_{l}(\hat{s})$ that are then used to obtain $\hat{F}_{\gamma}(x)$ in \eqref{SPA}.

Note that with the SPA approach, $\gamma_{l}$ do not have to follow the same fading distribution, i.e., the outage probability can be obtained, e.g.,  when Nakagami-$m$ and Hoyt (or any other combination of non-identical distributions) fading signals are combined at the receiver.

While \eqref{SPA} is a valid approximation for any value of the argument $x$, the expression can be further simplified noting that we are mostly interested in small values of the outage probability. In particular, assuming that $x<\mathbb{E}_{\gamma}[\gamma]$ implies that $\hat{s}<0$ (this is because we know that $\kappa''(s)
>0$ and $\kappa'(\hat{s})=0$). So, using the well known bound $\tr{erfc}(t) \leq \displaystyle{\frac{1}{t\sqrt{\pi}}\exp(-t^{2})}$ \cite[Appendix~II]{Martinez06}\footnote{Or, equivalently, $Q(t)\leq \phi(t)/t$. This bound holds for $t>0$ and becomes increasingly tight with increasing $t$.}, we replace $Q(-\hat{w})$ by $\displaystyle{\frac{-\phi(\hat{w})}{\hat{w}}}$, and obtain the following approximation of the CDF
\begin{align}\label{SPA.app}
   \hat{\tr{F}}_{\gamma}(x)\approx \tilde{\tr{F}}_{\gamma}(x)=-\frac{\phi(\hat{w})}{\hat{u}}=\frac{\tr{e}^{\kappa(\hat{s})-\hat{s}x}}{|\hat{s}|\sqrt{2\pi \kappa''(\hat{s})}}=\frac{\mc{M}(\hat{s})\tr{e}^{-\hat{s}x}}{|\hat{s}|\sqrt{2\pi \kappa''(\hat{s})}}.
\end{align}
We thus recover SPA forms similar to those used, for example, in \cite{Martinez06}.
 
\subsection{Identically Distributed Fading}

In the case of mutli-antenna receivers or ARQ transmissions, the channel between the transmitter and the receiver is essentially the same for each diversity ``branch''. Thus, it is reasonable to assume that $\gamma_{l}$ ($l=1,\ld, L$) are identically distributed. Then, $\kappa_{l}(s)\equiv\kappa_{1}(s)$ and the saddlepoint equation \eqref{SP.eq} reduces to
\begin{align}\label{SPA.L}
    L\cd\kappa'_{1}(\hat{s})=x.
\end{align}
In the cases of the fading distributions we consider, from Table~\ref{Table:CGF} we can easily see, that the solution of \eqref{SPA.L} is obtained solving the quadratic equation. These solutions are shows in Table~\ref{Table:SPA}. 

For Nakagami-$m$ fading, the solution is particularly simple and allows us to write the resulting approximation of the outage probability as 
\begin{align}
\tilde{\tr{F}}_{\gamma}(x)=
\frac{\left(x/L\ov{\gamma}\right)^{mL}}{\sqrt{2\pi mL}\left(1-x/L\ov{\gamma}\right)}\exp\bigl(-mL\cdot({x}/{L\ov{\gamma}}-1)\bigr),
\end{align}
while the exact solution is known in this case and given by
\begin{align}
\tr{F}_{\gamma}(x)= \Gamma\left( mL,\frac{mx}{\ov{\gamma}}\right)\!,
\end{align}
where $\Gamma(s,x)=\displaystyle{\frac{1}{\Gamma(s)}\int_0^x t^{s-1}\tr{e}^{-t}\tr{d}t}$ is the normalized lower incomplete gamma function and $\Gamma(s)=\Gamma(s,\infty)$.

\begin{table}[t]
\centering
\begin{tabular}{l||c|c}
Fading type & Parameter & $\hat{s}$\\
\hline\hline
Nakagami-$m$ &  $m$ & $\displaystyle{m\cd\left(\frac{1}{\ov{\gamma}}-\frac{L}{x}\right)}$ \\

Rice  &  $K $, $K'=K+1$ & $\displaystyle{\frac{K'}{\ov{\gamma}} -\frac{L}{2x}- \sqrt{{( \frac{K'}{\ov{\gamma}} -\frac{L}{2x} )}^{2}-{(\frac{K'}{\ov{\gamma}})}^{2}(\frac{x-L\ov{\gamma}}{x})}}$ \\
Hoyt  &   $q$, $q'={(q+1)}^{2}$ & $\displaystyle{\frac{q'}{4q\ov{\gamma}}-\frac{L}{2x}-\sqrt{(\frac{q'}{4q\ov{\gamma}}-\frac{L}{2x})^{2}  -\frac{q'(x-L\ov{\gamma})}{4xq{\ov{\gamma}^{2}}}}}$ 
\end{tabular}
\caption{Solution of the saddlepoint equation obtained combining $L$ signals with the same distribution.}\label{Table:SPA}
\end{table}

\subsection{Nakagami-$m$ Fading and $L=2$}

In the context of opportunistic relay-based communications, an equivalent three-terminal setup (source, selected relay, and destination) is quite common \cite{Benjillali10c}. In such a case, $\gamma_{1}$ (SNR between the source and the destination) and $\gamma_{2}$ (SNR between the relay and the destination) may have different average SNRs $\ov{\gamma}_{1}\neq\ov{\gamma}_{2}$ (e.g., when the distances between nodes are different). In this interesting scenario, we can also provide a closed-form solution when $\gamma_{1}, \gamma_{2}$ follow Nakagami-$m$ distributions. Equation \eqref{SP.eq} yields
\begin{align}\label{SP.eq.2}
\frac{m_{1}\ov{\gamma}_{1}}{m_{1}-\hat{s}\ov{\gamma}_{1}}+\frac{m_{2}\ov{\gamma}_{2}}{m_{2}-\hat{s}\ov{\gamma}_{2}}=x
\end{align}
which has the following solution
\begin{align}
\hat{s}=\ov{\beta}-\frac{\tilde{m}}{2x} -\sqrt{{(\ov{\beta}-\frac{\tilde{m}}{2x})}^{2}-\frac{\hat{m}(x-\tilde{\gamma})} {x\hat{\gamma}}}
\end{align}
with $\tilde{\gamma}=\ov{\gamma}_{1}+\ov{\gamma}_{2}$, $\hat{\gamma}=\ov{\gamma}_{1}\cd \ov{\gamma}_{2}$, $\tilde{m}= m_{1}+ m_{2}$, $\hat{m}= m_{1}\cd m_{2}$, and $\ov{\beta}=(m_{1}/\ov{\gamma}_{1}+m_{2}/\ov{\gamma}_{2})/2$.

\subsection{General Case}

In the general case, we are not able to solve \eqref{SP.eq} in closed-form. However, the solution of $\kappa'(s)=x$ may be found in a few recursive steps which implement the Newton method for solving non-linear equations
\begin{align}\label{Newton}
\hat{s}_{k}&=\hat{s}_{k-1}+  \frac{x-\kappa'(\hat{s}_{k-1})}{\kappa''(\hat{s}_{k-1})},\qquad \text{for}\quad k=1,\ld, K_\tr{max}
\end{align}
where $\hat{s}_{0}$ is the initial solution, and the number of recursive steps, $K_\tr{max}$, controls the accuracy of the solution. For the numerical examples we will discuss later, a satisfactory accuracy of the approximation is obtained with a relatively small $K_\tr{max}<5$.

While the requirement of finding $\hat{s}$ recursively is a drawback of the SPA-based method compared to closed-form solutions shown before, we note that the existing alternatives often imply a similar computational effort. For example, finding the exact form of the CDF $\tr{F}_{\gamma}(x)$ for Nakagami-$m$ variables calls for finding infinite series of coefficients defined recursively \cite{Alouini01},  a numerical integration implies a summation over the integration nodes \cite{Ko00}, and  approximations shown in \cite{daCosta08} require solving non-linear equations. Thus, from the complexity point of view, SPA is a simple approach to calculate the outage.

\section{Numerical Examples}
In this section, we present a few numerical results to assess the accuracy of the proposed SPA-based evaluation of the outage probability.

\begin{figure}[tb]
\psfrag{xlabel}[c][c][1.2]{$x$~[dB]}
\psfrag{ylabel}[c][c][1.2]{CDF}
\psfrag{Exact}[l][l][1.2]{$F_{\gamma}(x)$}
\psfrag{Saddlepoint}[l][l][1.2]{$\hat{F}_{\gamma}(x)$}
\psfrag{simple}[l][l][1.2]{$\tilde{F}_{\gamma}(x)$}
\psfrag{Nakagami--m}[l][l][1.2]{Nakagami-$m$}
\psfrag{Rice}[l][l][1.2]{Rice}
\psfrag{Hoyt}[l][l][1.2]{Hoyt}
\psfrag{L=2}[l][l][1.2]{\hskip -0.25cm $L{=}2$}
\psfrag{L=5}[l][l][1.2]{\hskip -0.2cm $L{=}5$}
\psfrag{L=8}[l][l][1.2]{\hskip -0.1cm $L{=}8$}
\begin{center}
\scalebox{0.7}{\includegraphics{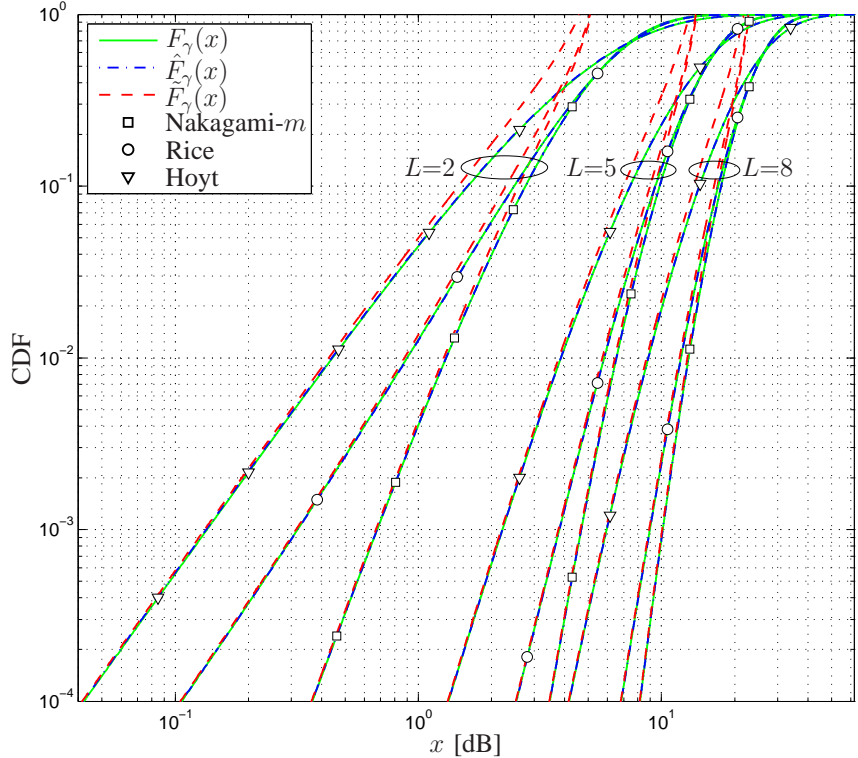}}
\caption{Exact CDF $F_{\gamma}(x)$ compared with the SPA $\hat{F}_{\gamma}(x)$ and the simplified SPA $\tilde{F}_{\gamma}(x)$ when $\gamma$ is the aggregate SNR from $L$ diversity branches with $\ov{\gamma}_{l}=5$dB and underling distributions: Nakagami-$m$ ($m_{l}=2$), Rice ($K_{l}=2$), and Hoyt ($q_{l}=0.5$).}\label{Fig:Liid}
\end{center}
\end{figure}

In Fig.~\ref{Fig:Liid}, we show the results when $\gamma$ is the sum of $L$ identically distributed $\gamma_{l}$, each with $\ov{\gamma}_{l}=5$dB. Three cases are presented: a) Nakagami-$m$ fading signals with $m_{l}=2$, b) Rice fading with $K_{l}=2$, c) Hoyt fading with $q_{l}=0.5$.

\begin{figure}[tb]
\psfrag{xlabel}[c][c][1.2]{$x$~[dB]}
\psfrag{ylabel}[c][c][1.2]{CDF}
\psfrag{exact}[l][l][1.2]{$F_{\gamma}(x)$}
\psfrag{saddle}[l][l][1.2]{$\hat{F}_{\gamma}(x)$}
\psfrag{simple}[l][l][1.2]{$\tilde{F}_{\gamma}(x)$}
\psfrag{Y=5dB}[c][c][1.2]{$\ov{\gamma}_{1}=5$dB}
\psfrag{Y=8dB}[c][c][1.2]{~~$\ov{\gamma}_{1}=8$dB}
\psfrag{Y=12dB}[c][c][1.2]{~~$\ov{\gamma}_{1}=12$dB}
\begin{center}
\scalebox{0.7}{\includegraphics{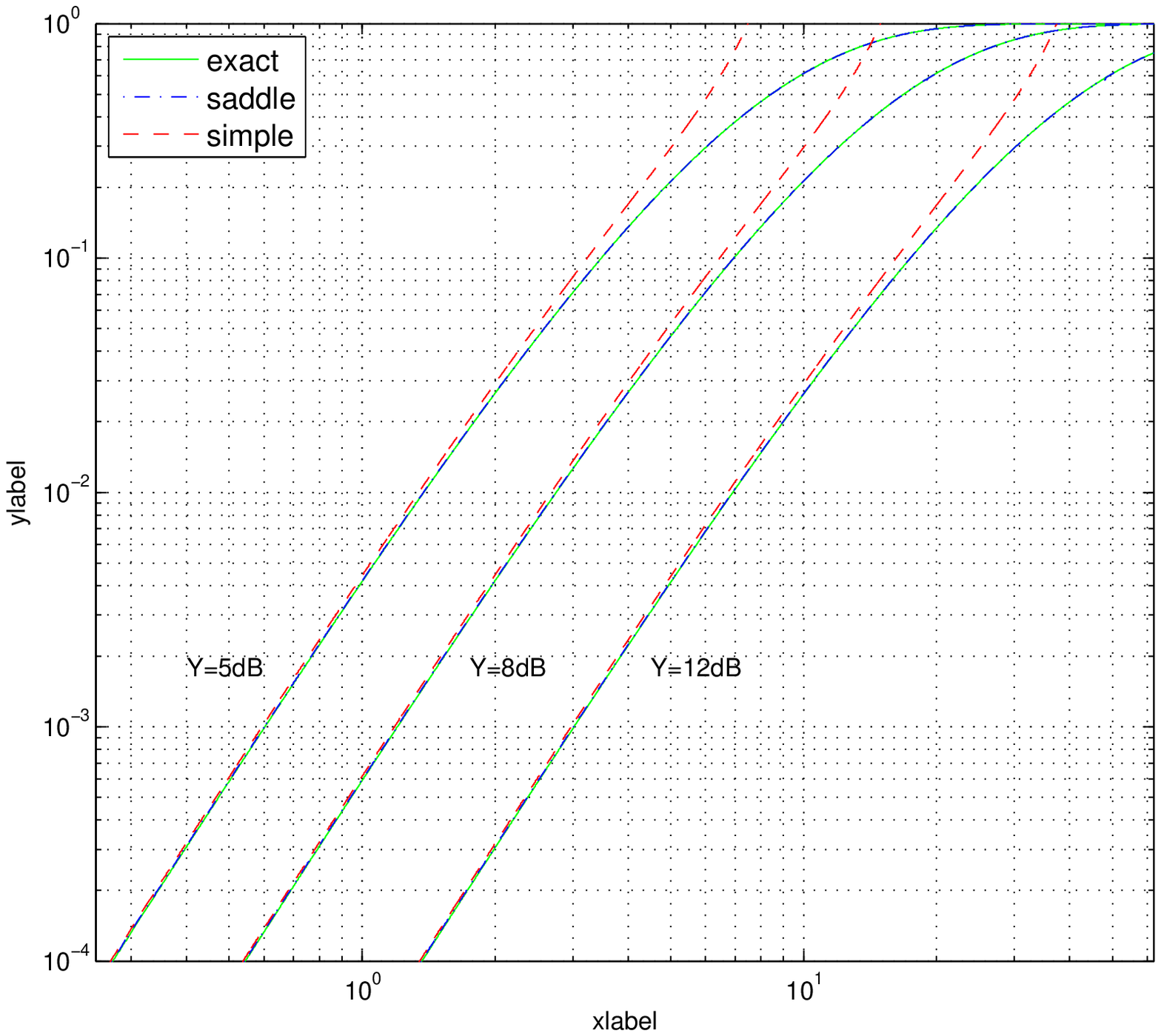}}
\vskip -0.2cm
a)

~
\vskip 0.2cm

\scalebox{0.7}{\includegraphics{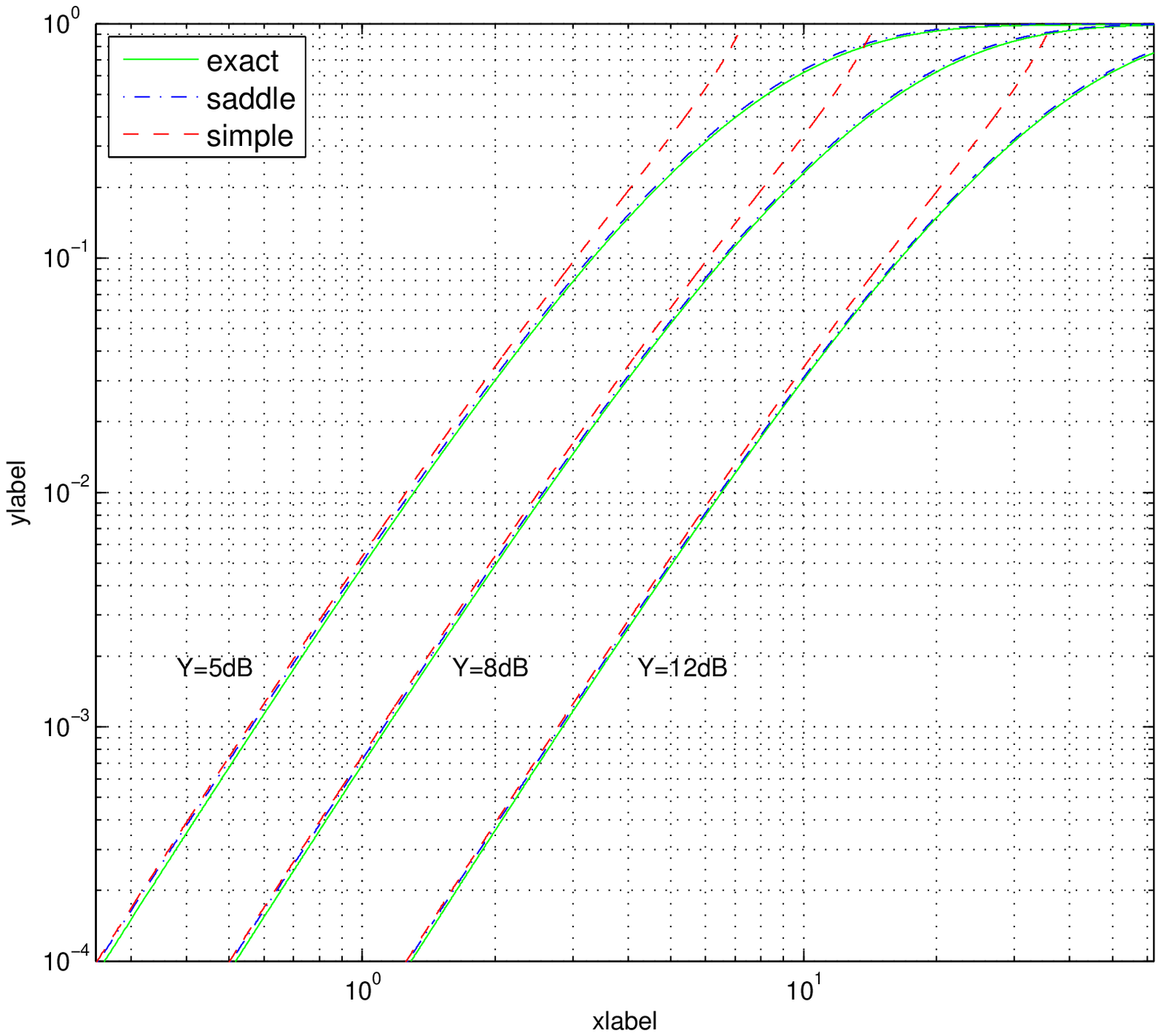}}
\vskip -0.2cm
b)

\caption{Exact CDF $F_{\gamma}(x)$ is compared with the SPA $\hat{F}_{\gamma}(x)$ and the simplified SPA $\tilde{F}_{\gamma}(x)$ when $L=2$ and both variables correspond to Nakagami-$m$ fading with a)~$m_{1}=1$ and $m_{2}=2$, and b)~$m_{1}=0.5$ and $m_{2}=2.5$; $\ov{\gamma}_{2}=2\ov{\gamma}_{1}$.}\label{Fig:L2}
\end{center}
\end{figure}

In Fig.~\ref{Fig:L2}, we show the results for Nakagami-$m$ distributed signals with $L=2$, $\gamma_{1}=5, 8, 12$dB and $\ov{\gamma}_{2}=2\ov{\gamma}_{1}$. We consider two cases for the shape parameters: a) $m_{1}=1$ and $m_{2}=2$, and b) $m_{1}=0.5$ and $m_{2}=2.5$. Case (a) can also be solved using the closed-form expressions from \cite{Benjillali10c}, while case (b) requires expressions given in \cite{Alouini01} with a truncation of the infinite series. Our method solves both cases (a) and (b) in closed-form. 

\begin{figure}[tb]
\psfrag{xlabel}[c][c][1.2]{$x$~[dB]}
\psfrag{ylabel}[c][c][1.2]{CDF}
\psfrag{exact}[l][l][1.2]{$F_{\gamma}(x)$}
\psfrag{saddle}[l][l][1.2]{$\hat{F}_{\gamma}(x)$}
\psfrag{simple}[l][l][1.2]{$\tilde{F}_{\gamma}(x)$}
\psfrag{Y=5dB}[c][c][1.2]{$\ov{\gamma}_{1}=5$dB}
\psfrag{Y=8dB}[c][c][1.2]{~~~$\ov{\gamma}_{1}=8$dB}
\psfrag{Y=12dB}[c][c][1.2]{~~~$\ov{\gamma}_{1}=12$dB}
\begin{center}
\scalebox{0.7}{\includegraphics{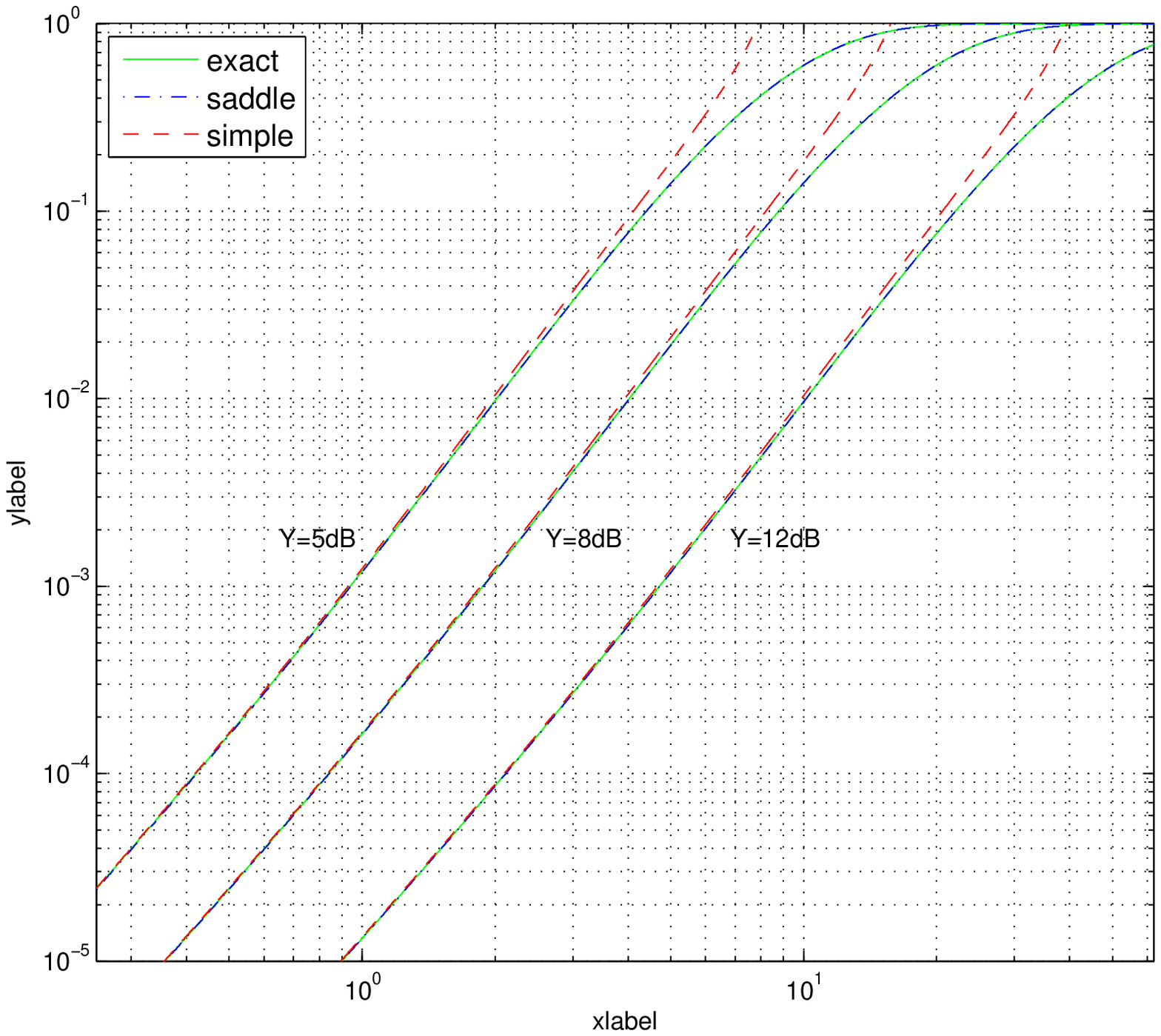}}

\caption{Exact CDF $F_{\gamma}(x)$ is compared with the SPA $\hat{F}_{\gamma}(x)$ and the simplified SPA $\tilde{F}_{\gamma}(x)$ for $L=2$ when one variable corresponds to Nakagami-$m$ fading with $m=1.5$  and another -- to the Rice fading with $K=5$; $\ov{\gamma}_{2}=2\ov{\gamma}_{1}$.}\label{Fig:NR}
\end{center}
\end{figure}

Finally, we show in Fig.~\ref{Fig:NR} the outage results after combining $L=2$ differently fading signals corresponding to a Nakagami-$m$ fading with parameters $m=1.5$, $\ov{\gamma}_{1}=5, 8, 12$dB and a Rice fading with $K=5$ and $\ov{\gamma}_{2}=2\ov{\gamma}_{1}$. The saddlepoint solution was solved using \eqref{Newton} with $K_\tr{max}=5$.

These examples illustrate well how exact the SPA method is, yielding results that are practically identical to the exact form of the CDF. The simplification \eqref{SPA.app} also provides very accurate results for all outage values below $10^{-1}$ which, in most cases will be the region of interest when evaluating the performance of practical diversity combining schemes.

\section{Conclusions}\label{Sec:Conclusions}

In this work we proposed to use a saddlepoint approximation to evaluate the outage probability at receivers combining signals from arbitrarily drawn distributions. We have shown that knowing the corresponding moment generating functions, we are able to accurately estimate the outage for an arbitrary number of combined signals. The solution requires, in general, solving a scalar non-linear equation whose solution can even be found, in particular cases, in closed-form.


\bibliography{IEEEabrv,references_all}
\bibliographystyle{IEEEtran}

\end{document}